# BRAIN-COMPUTER INTERFACES AND QUANTUM ROBOTS

Eliano Pessa
*Centro Interdipartimentale di Scienze Cognitive, Università di Pavia.*
*Piazza Botta 6, 27100 Pavia, Italy.*
*eliano.pessa@unipv.it*

Paola Zizzi
*Dipartimento di Matematica Pura e Applicata, Università di Padova.*
*Via Trieste, 63, 35121 Padova, Italy.*
*zizzi@math.unipd.it*

**Abstract**

The actual (classical) Brain-Computer Interface attempts to use brain signals to drive suitable actuators performing the actions corresponding to subject's intention. However this goal is not fully reached, and when BCI works, it does only in particular situations. The reason of this unsatisfactory result is that intention cannot be conceived simply as a set of classical input-output relationships. It is therefore necessary to resort to quantum theory, allowing the occurrence of stable coherence phenomena, in turn underlying high-level mental processes such as intentions and strategies. More precisely, within the context of a dissipative Quantum Field Theory of brain operation it is possible to introduce generalized coherent states associated, within the framework of logic, to the assertions of a quantum metalanguage. The latter controls the quantum-mechanical computing corresponding to standard mental operation. It thus become possible to conceive a Quantum Cyborg in which a human mind controls, through a quantum metalanguage, the operation of an artificial quantum computer.

## 1. Introduction

The Brain-Computer Interfaces (BCI) are systems that acquire and analyze brain signals (typically of electromagnetic nature) to create high-bandwidth communication channels in real time between the human brain and a computer (see for an overview, e.g., Dornhege *et al*., 2007). Most often BCI are designed to capture subject's intentions in order to drive suitable actuators, performing the actions wanted by the subject himself. However, even if BCI seem to open the way for a deep merging between human minds and computers, their actual implementations still appear as unsatisfying and very far from reaching the goal of a complete integration between human beings and artificial devices.

In this paper we will introduce some arguments supporting the impossibility of reaching this goal within the design framework actually used for BCI. In short, this impossibility is due to the fact that actual BCI are designed to allow a *computer-to-computer* communication within a classical context. However, recent studies lead to the conclusion that the human mind is *not* a classical computer, and, in general, not completely reducible to any kind of computer (not even classical) because of the non-algorithmic nature of some mental processes. Moreover, we will argue that most mental processes should be described by Quantum Physics. Among these processes there are control processes acting on most mental operations which, otherwise, could not be performed. As control processes can be seen as a sort of metathought, the logic underlying them can be viewed as the one of a (quantum) *metalanguage* describing most high-level mental processing, such as reasoning, decision making, recalling from memory, and the like. A quantum metalanguage reflects into a quantum object language, and controls the latter.

If we adopt this theoretical framework, it follows that, provided we have at disposal new kinds of BCI, allowing a *quantum-computer-to-quantum-computer* communication, we could use human mind to control, through its quantum metalanguage, the operation of an artificial quantum computer. The whole system constituted by a human subject and an artificial quantum computer (controlled by the quantum metalanguage of the subject himself) is a new kind of cyborg, called *Quantum Cyborg* (QC). The latter (see Zizzi, 2008) would allow a deeper integration between human mind and artificial devices. As the practical implementation of a QC requires to solve a number of difficult problems, such as the one of avoiding decoherence induced by external environment, in the final part of this paper we will shortly outline some possible strategies for coping with these difficulties.

## 2. The conceptual problems underlying BCI design

The design of a BCI requires the solution of a number of hard problems:
1) Knowing what features of brain signals are associated to specific kinds of intentions or of mental states.
2) Selecting the best techniques to detect these features in presence or noise and artefacts.
3) Finding the best way to implement online the sequence detection-action performance.

So far these problems have been dealt with by resorting, on one hand, to experiments on human subjects which imagine to perform a given action, and, on the other hand, to soft computing algorithms, like the ones allowing artificial neural networks to learn, relying only on examples, to mimic whatever kind of input-output relationship (see, in this regard, Bishop, 1995; Rojas, 1996; Bartlett and Anthony, 1999).

This strategy, of course, is based on the hypothesis that mental states are fully characterized by specific activation patterns of brain neurons, where the attribute 'specific' is to be understood in a functional sense. This means that, while the same mental state is not necessarily associated to the activation of the same brain neurons, what matters is that, whatever be the neurons activated, each time, in correspondence to this state, they give rise to the same input-output behavioural patterns of the subject lying in this mental state itself.

Within this functionalist framework the problem arising from the fact that the same mental state is, each time, associated to different patterns of neural activation is avoided by supposing that all these different patterns are characterized, if associated to the same mental state, by a sort of invariant "signature". The latter can be conceived as defined by a set of invariant features characterizing these patterns, and whence also the electromagnetic signals emitted by the brain and detected, for instance, through electroencephalograms (EEG). These features, in principle, could be found through a suitable analysis of the observed EEG. However, as this analysis should, with high probability, be very difficult to implement, the best strategy seems to be the one of collecting the highest number possible of associations between EEG signals and motor outputs, so as to train, through a supervised learning procedure based on known examples, an artificial neural network to produce the output only when the presented EEG corresponds to the intention. Thus, if the learning would be successful, the network weight distribution found after the completion of learning itself, would automatically give an implicit description of the procedure to be used to analyze the EEG signal to extract the 'signature' of the intention.

Without entering into technical details about the implementation of this strategy, widely used to design the actual BCI, we will limit ourselves to remark that it is, in principle, destined to fail, owing to the existence of two main conceptual difficulties. The first one stems from the fact that, given a whatever supervised learning procedure, its performance in the test phase (that is, after completing the initial training phase) depends in a crucial way on the training examples used during the learning. Namely, not only their number must be high enough, but we also need that they are representative of the possible kinds of situations occurring within the whole sample from which training examples themselves have been extracted. Unfortunately, both conditions will never be satisfied in the case of EEG signals (or of whatever other kind of brain signal), first because the number of available data is severely limited (for practical reasons connected to the way through which experiments on human subjects are performed), and, in the second place, because we do not know (and we will never know) how the whole sample of possible EEG is structured. The latter circumstance precludes any possibility of assessing the representatives of chosen training examples, so that, for every supervised learning procedure, we will always be unable to grant its reliability.

However, even if this difficulty could be avoided, we could never overcome the second main conceptual difficulty, arising from the fact that, in principle, mental states cannot be defined only in terms of input-output associations. The number of different possible input-output relationships associated to the intention is virtually unlimited, just because the number of different possible contexts is unlimited. We stress here that the word 'context' includes not only states of the environment, but also the occurrence of other, contemporarily present, mental states. To conclude this section, the previous arguments show that the actual strategy used to design BCI is unsuited to capture the occurrence of intentional states in the minds of subjects. This means that the main goal underlying the introduction of BCI will never be reached in this way. In the next sections we will explore a possible alternative.

## 3. Metathought

Before going further, we remark that intentions can be hardly conceived as mental states. Namely the concept itself of mental state is useless when dealing with adaptive mechanisms, such as intentions, which underlie a number of control processes, in turn acting on mental operations, such as reasoning, deciding, recalling, and the like. If we generically denote the whole set of usual mental operations through the word 'thought', we should denote intentions, and other control mechanisms, through the word 'metathought', to stress the fact the latter acts on, controls, and drives the ordinary thought.

Within the history of Psychology the concept of metathought has not been very popular. In the Seventies Flavell introduced an analogous concept under the name 'metacognition' (Flavell, 1976). The study of these kinds of topics has been pursued mostly within the domain of Developmental and Educational Psychology (see, for instance, Weinert and Kluwe. 1987; Crowley *et al.*, 1997). Only in more recent times some authors began to introduce computational models of the operation of prefrontal cortex, considered as the seat of control processes within the brain (Becker and Lim, 2003). Anyway, all metathought processes could be interpreted as aiming to keep some sort of equilibrium or, more in general, of coherence. Therefore, in order to describe metathought and intentions, the problem is to find what are the best models, as regards both the physical basis of these processes and their logical nature.

Of course, the generally adopted solution of this problem consists in resorting to classical physics and to classical logic. Unfortunately both are ruled out by theoretical arguments as well as experimental findings. On the theoretical side, we know from long time that classical physics is not endowed with coherence-keeping mechanisms. The latter are forbidden by Second Principle of Thermodynamics or, what is equivalent, by the so-called Correlation Weakening Principle, stating that whatever long range correlation will die away after a long enough evolution time. On the experimental side, a large number of experiments performed by psychologists evidenced that most mental processes, including semantic memory search, problem solving, reasoning, cannot be described by classical logic, which, rather, appears to be more suited to describe the operations performed on bits within a digital computer (see Adler and Rips, 2008).

On the contrary, Quantum Theory appears as endowed with powerful coherence-keeping mechanisms, whose efficiency is, in rough terms, due to the fact that within it whatever entity is not spatially and temporally localized but rather described by a probability distribution ranging over the whole space-time. Thus, the superposition of different probability distributions associated to different entities gives rise to a sort of long range correlation between these latter which counteracts the disturbing influences produced by heat, noise, and other coherence-destroying mechanisms. Among the coherence phenomena of quantum nature we can quote ferromagnetism, super fluidity, laser effect, superconductivity, and many others. Some of the latter occur only below a very low critical temperature, close to the absolute zero, but others take place even at high temperature.

It is, however, to be remembered that the expression 'Quantum Theory' is too generic. Namely we currently have two different kinds of Quantum Theories: *Quantum Mechanics* (QM), dealing with fixed numbers of particles lying within finite space volumes, and *Quantum Field Theory* (QFT), in which the field strengths are the basic entities, and infinite volumes as well as processes of creation and destruction of particles are possible. While in QM we have a finite number of degrees of freedom, QFT is characterized by an infinite (and continuous) number of degrees of freedom. Both in QM and in QFT the mathematical entities describing physical quantities must fulfil suitable constraints, expressing the non-classical nature of these theories and often called *canonical commutation relations* (CCR). Once given a physical system, a particular choice of the description of its dynamics, provided it fulfils the CCR, is called a *representation* of the CCR. Now an important theorem of QM, proved many years ago by Von Neumann, states that within it all different representations of the same physical system are unitarily equivalent. This means that in QM all representations of a given physical system have the same *physical content*. However, this no longer true in QFT, as shown already in the Sixties. This circumstance entails that within the latter theory the different descriptions of the same physical system can be unitarily non-equivalent, that is describing *different kinds of physics*. Such a state of affairs occur just when we deal with *phase transitions*, when a given physical system can undergo a transformation from a given phase (for instance solid) to another phase (for instance liquid), the two phases being characterized by entirely different physical properties. This implies that only QFT offers a framework for describing phase transitions (see on these topics Minati and Pessa, 2006, Chap. 5; Pessa, 2008).

These considerations entail that only QFT can describe the emergence of metathought. Such a circumstance is at the basis of a number of *Quantum Brain Theories* (see, for comprehensive overviews, Jibu and Yasue, 1995; Vitiello, 2001; Globus *et al.*, 2004), in turn relying on a firm experimental evidence about the quantum nature of physical phenomena underlying mental processes (Tuszyński, 2006; Abbott *et al.*, 2008). We cannot, however, forget that the same evidence leads us to conclude that normal thought processes should be described by QM, the theory to which QFT is reduced when we are far from phase transitions and the number of components of our systems is kept constant. The quantum logic of mind [Zizzi, PhD thesis] describes a *Quantum Computation* acting on *qubits*, entities consisting in a superposition of two quantum states, conventionally denoted as '0' and '1'. Each qubit can be seen as carrying a sort of implicit double potentiality, which can give rise to an ordinary bit under the action of a projection operator producing a collapse of the qubit state. Thus the normal operation of human mind, in a number of cases, can be viewed as equivalent to the one of a suitable *Quantum Computer* manipulating qubits.

**4. Quantum Robots**

The previous considerations lead in a natural way to the introduction of the concept of *Quantum Robot* (QR). The latter, first proposed by Benioff (cfr. Benioff, 1998), can be defined as a mobile system which has a quantum computer on board, and any needed ancillary systems. A QR moves in and interacts with the environment of a quantum system. However, the QR originally discussed by Benioff have no awareness of their environment, and do not make decisions or measurements. We can therefore ask ourselves whether in the future it might be that quantum robots will be aware of the environment, and could perform experiments. This means that they might even become self-aware, conscious, and have "free will".

In this regard it is to be taken into account that, in order to endow a QR with these features, it should be equipped with a sort of "internal observer" able both to look at the internal (quantum) computations of the QR itself and to control them. This internal observer thus should act on QR computations through a *Quantum Metalanguage* suited to control a *Quantum Language*, expressed in terms of qubit manipulations. It is natural to suppose that the core of the Quantum Metalanguage consists of inner measurement operators, like the ones used in QM. Unfortunately the latter are not suited to control a QR, as they coincide with projection operators, destroying qubits (which, after the action of a projector, become simple classical bits, whose value is 0 or 1). This entails that traditional "Quantum Logic" would not correctly describe the inner measurements needed by a QR (Zizzi, 2007).

A possible way out of this problem consists in resorting to the so-called "Weak Measurements" (WM), yet introduced by Aharonov *et al.* (see Aharonov *et al.*, 1988). Without entering here into technical details, we will limit ourselves to mention that a WM is based on a measuring apparatus which interacts very weakly with the quantum system to be measured so as to introduce in it only a negligible perturbation. Moreover, after the interaction with the measure apparatus, the latter acts in such a way as to measure (this time in a projective way), not the physical quantity which is the goal of the measure itself, but another different physical quantity, characterizing the system's environment. The result of the latter measure, however, allows to guess the searched value of the physical quantity characterizing the system under study, value which was the true goal of the measurement procedure. A conceptual analysis, which will not be reported here (see Zizzi, 2005; 2006), leads to represent WM through non-hermitian operators, whose eigenvalues (the values of the measured quantities) are given, in general, by complex numbers, with a real and an imaginary part. These operators can be interpreted as describing a quantum system interacting with a *dissipative environment* (see Vitiello, 2001; Pessa, 2008). In this regard we recall that the physical processes occurring within the brain should be dealt with, at least as concerns quantum aspects, through a *Dissipative* QFT, in absence of which the brain dynamics would be characterized by only a single ground state, rather that by the multiplicity of different ground states needed to accommodate the multitude of different memory states necessarily occurring within the brain.

It is to be stressed that a control based on a Quantum Metalanguage made by non-hermitian operators solves in an easy way the problem of the *decoherence* of QR. The latter, as it is well known, is produced by the action of a *thermal environment* destroying the superpositions of quantum states. However, as the non-hermitian operators associated to WM describe an open system interacting with a *dissipative environment* (not coincident with the thermal one), we must take into account that the very existence of qubits results from an entanglement between the system and this environment. Thus, if the states of the dissipative environment are eigenstates of these operators (remember that WM act on the dissipative environment), every thermal perturbation acting on the system will be automatically counteracted by the entanglement of system's eigenstates with the ones of dissipative environment, which will resist against any attempt to entangle the system itself with the thermal environment.

All previous arguments points to an interesting possibility, the one of using the generalized coherent states (eigenstates of non-hermitian operators) of the brain to control a quantum system, for instance a quantum computer based on quantum dots. This would open the way to the implementation of a QC, in which a human subject, through the quantum metalanguage, could drive a QC, through a BCI much more powerful than the actually existing ones, and able to transform in a more effective way human intentions into actions. Such a kind of QC, opening the way to a deeper merging of humans and computers (possible owing to the quantum framework), would, however, require a lot of experiments, and conceptual as well as technological advances. While, in principle, nobody prevents from having a quantum robots, endowed with QFT-based aspects, able to perform inner WM on its own quantum operations, it seems more plausible that the only feasible implementation of a Quantum Metalanguage be the one based on human brain.

## 5. Conclusions

The previous arguments showed that the quantum approach predicts the possibility of a direct action of mind on matter. This circumstance, beyond the improvement of the operation of existing BCI, opens the possibility of designing new kinds of BCI interfaces. This could cause a revolutionary change in our actual way of thinking, based on the tacit assumption that our thoughts have no direct effects on the world.

**References**

Abbott, D., Davies, P.C.W., Pati, A.K. (Eds.) (2008). *Quantum aspects of life*. Imperial College Press, London.

Adler, J.E., Rips, L.J. (Eds.) (2008). *Reasoning. Studies of human inference and its foundations*. Cambridge University Press, New York.

Aharonov, Y., Albert, D.Z., Vaidman, L. (1988). How the result of a measurement of a component of the spin of a spin-1/2 particle can turn out to be 100. *Physical Review Letters*, **60**, 1351-1354.

Bartlett, P.L., Anthony, M.M. (1999). *Neural Network Learning: Theoretical Foundations*. Cambridge University Press, Cambridge, UK.

Becker, S., Lim, J. (2003). A computational model of prefrontal control in free recall: Strategic memory use in the California verbal learning task. *Journal of Cognitive Neuroscience*, **15**, 821-832.

Benioff, P. (1998). Quantum Robots and environments. *Physical Review A*, **58**, 893-904.

Bishop, C.M. (1995). *Neural networks for pattern recognition*. Oxford University Press, Oxford, UK.

Crowley, K., Shrager, J., Siegler, R. S. (1997). Strategy Discovery as a Competitive Negotiation between Metacognitive and Associative Mechanisms. *Developmental Review,* **17**, 462-489.

Dornhege, G., Millán, J. del R., Hintenberger, T., McFarland, D., Müller K.-R. (Eds.) (2007). *Towards Brain-Computer Interfacing*. MIT Press, Cambridge, MA.

Flavell, J. H. (1976). Metacognitive Aspects of Problem Solving. In L.Resnick (Ed.). *The Nature of Intelligence*. (pp. 231-235). Erlbaum, Hillsdale, NJ.
Globus, G.G., Pribram, K.H., Vitiello, G. (Eds.) (2004). *Brain and being. At the boundary between science, philosophy, language and arts*. Benjamins, Amsterdam.
Jibu, M., Yasue, K. (1995). *Quantum Brain Dynamics and Consciousness: An Introduction*. Benjamins, Amsterdam.
Minati, G., Pessa, E. (2006). *Collective Beings*. Springer, Berlin.
Pessa, E. (2008). Phase Transitions in Biological Matter. In I.Licata, A.Sakaji (Eds.), *Physics of Emrgence and Organization* (pp. 165-228). World Scientific, Singapore.
Rojas, R. (1996). *Neural networks. A systematic introduction.* Springer, Berlin-Heidelberg-New York.
Tuszyński, J.A. (Ed.) (2006). *The emerging physics of consciousness*. Springer, Berlin.
Vitiello, G. (2001). *My double unveiled*. Benjamins, Amsterdam.
Weinert, F.E., Kluwe, R.H. (Eds.) (1987). *Metacognition, motivation, and understanding*. Erlbaum, Hillsdale, NJ.
Zizzi, P. (2005). Qubits and quantum spaces. *International Journal of Quantum Information*, **3**, 287-291.
Zizzi, P. (2006). Theoretical setting of inner reversible quantum measurements. *Modern Physics Letters A*, **21**, 2717-2727.
Zizzi, P. (2007). Basic Logic and Quantum Entanglement. *Journal of Physics Conf. Ser.*, **67**, 012045.
Zizzi, P. (2008). "I, Quantum Robot: Quantum Mind Control on a Quantum Computer" arXiv: 0812.4614.